\documentclass{jfm}
\usepackage{graphicx}
\usepackage{epstopdf, epsfig}
\usepackage{color}
\usepackage{amsmath}

\usepackage{amssymb}
\usepackage{ulem}

\usepackage{scalerel,stackengine}
\stackMath
\newcommand\widecheck[1]{%
\savestack{\tmpbox}{\stretchto{%
  \scaleto{%
    \scalerel*[\widthof{\ensuremath{#1}}]{\kern-.6pt\bigwedge\kern-.6pt}%
    {\rule[-\textheight/2]{1ex}{\textheight}}
  }{\textheight}%
}{0.5ex}}%
\stackon[1pt]{#1}{\scalebox{-1}{\tmpbox}}%
}

\newcommand{\bxs}{\boldsymbol{x_s}}

\newcommand{\dbar}{\mathchar'26\mkern-8mu \delta}

\shorttitle{Multi-scale statistics of active matter turbulence}
\shortauthor{J. Urzay, A. Doostmohammadi and J.~M. Yeomans}

\title{Multi-scale statistics of turbulence motorized by active matter}

\author{J. Urzay,\aff{1}
  \corresp{\email{jurzay@stanford.edu}}$\ddagger$
 A. Doostmohammadi\aff{2}\corresp{Both authors contributed equally to this work}~
   \and J.~M. Yeomans\aff{2}}

\affiliation{\aff{1}Center for Turbulence Research, Stanford University, CA 94305-3024, USA
\aff{2}Rudolf Peierls Centre for Theoretical Physics, University of Oxford, OX1 3NP, UK
}

\begin{document}

\maketitle

\begin{abstract}
A number of micro-scale biological flows are characterized by spatio-temporal chaos. These
include dense suspensions of swimming bacteria, microtubule bundles driven by motor proteins, and dividing and migrating confluent layers of cells. A characteristic common to all of these systems is that they are laden with active matter, which transforms free energy in the fluid into kinetic energy. Because of collective effects, the active matter induces multi-scale flow motions that bear strong visual resemblance to turbulence. In this study, multi-scale statistical tools are employed to analyze direct numerical simulations (DNS) of periodic two- (2D) and three-dimensional (3D) active flows and compare them to classic turbulent flows. Statistical descriptions of the flows and their variations with activity levels are provided in physical and spectral spaces. A scale-dependent intermittency analysis is performed using wavelets. The results demonstrate fundamental differences between active and high-Reynolds number turbulence; for instance, the intermittency is smaller and less energetic in active flows, and the work of the active stress is spectrally exerted near the integral scales and dissipated mostly locally by viscosity, with convection playing a minor role in momentum transport across scales.

\end{abstract}

%
\section{Introduction}
The multi-scale processes observed in the types of flows discussed here are induced by active matter laden in a fluid \citep{Julia2012,Dogic2012,Jorn2013}. These are a special class of multi-phase flows, where the constituent particles are self-propelled. Examples of biological active matter are cells, motor proteins and bacteria. Synthetic active matter can be manufactured in the form of mechanically, chemically or optically propelled particles. However, a unifying characteristic of laden active matter is that it transforms free energy in the fluid into systematic motion \citep{Sriram2002}. Although such energy conversion occurs at the particle scales, the collective interactions among many of these particles oftentimes translate into unstable flow motion across much larger scales. 

Despite the low Reynolds numbers involved, flows induced by active matter have been referred to as {\em active turbulence} in analogy to the unsteady multi-scale dynamics found in high-Reynolds number flows \citep{Julia2012,Frey2015,Giomi2015}. However, whether these flows can be identified as turbulent in the classical sense is debatable. The conservation equations of active flows are considerably non-linear and formally more complex than the incompressible Navier-Stokes equations, and the nature of these non-linearities is fundamentally different from those responsible for high-Reynolds number turbulence. Non-linear dynamical systems are known to display chaotic particle trajectories and mixing behavior even at low Reynolds numbers \citep{Ottino}.

In this study multi-scale tools are employed to examine basic flow statistics in 2D and 3D. The remainder of this paper is organized as follows. The formulation and computational set-up are described in \S\ref{formulation}. The analysis of the numerical results is presented in \S\ref{qual}. Finally, concluding remarks are provided in \S\ref{conclusion}. 
~\\[-0.3in]
\section{\label{formulation}Formulation and computational setup}
The approach employed here is based on the numerical integration of the conservation equations of dense active nematohydrodynamics. This continuum formulation extends the description of passive liquid crystals of \cite{DeGennesBook} and has proven successful in reproducing spatio-temporal dynamics observed in experiments of active flows (see \cite{ourNComm2016} and references therein).

\subsection{Conservation equations for active nematohydrodynamics}
In this formulation, the mesoscopic orientational order of active particles is represented by the nematic tensor $Q_{ij}=(3q/2)(n_in_j-\delta_{ij}/3)$, where $q$ is the magnitude of the nematic order, $n_i$ is the director and $\delta_{ij}$ is the Kronecker delta. The conservation equation for the nematic tensor is given by
\begin{equation}
\partial_{t} Q_{ij} + u_k \partial_{x_k} Q_{ij}=  \Gamma H_{ij}+R_{ij},
\label{eqn:lc}\\
\end{equation}
where $u_k$ are velocity components, $t$ is time, $x_i$ are spatial coordinates, and $\Gamma$ is proportional to a rotational diffusivity. Additionally, $R_{ij}$ is a standard co-rotation tensor defined as 
\begin{equation}
R_{ij} = \left(\lambda S_{ik} + \Omega_{ik}\right)\left(Q_{kj} + \frac{\delta_{kj}}{3}\right) + \left(Q_{ik} + \frac{\delta_{ik}}{3}\right)\left(\lambda S_{kj} - \Omega_{kj}\right) - 2 \lambda \left(Q_{ij} + \frac{\delta_{ij}}{3}\right)Q_{kl}\partial_{x_k} u_l,
 \label{eqn:cor}
\end{equation}
which accounts for the response of the orientational field to the extensional and rotational components of the velocity gradients, with $S_{ij} = (1/2)(\partial_{x_i} u_j + \partial_{x_j} u_i)$ and $\Omega_{ij} = (1/2)(\partial_{x_j} u_i - \partial_{x_i} u_j)$ being the strain-rate and vorticity tensors, respectively. The relative importance of the vorticity and strain rate is controlled by the  parameter $\lambda$, which characterizes the alignment of the nematics with the flow. The term involving  $\Gamma$ in (\ref{eqn:lc}) describes, through the auxiliary tensor $H_{ij}= -\dbar \mathcal{F}/\dbar Q_{ij} + (\delta_{ij}/3) {\rm tr} (\dbar \mathcal{F}/\dbar Q_{kl})$, the relaxation of $Q_{ij}$ to a minimum of a free energy
 \begin{equation}
 \mathcal{F}=  (A/2)Q_{ij} Q_{ij} + (B/3) Q_{ij} Q_{jk} Q_{ki}+ (C/4)(Q_{ij} Q_{ij})^2+ K/2\left( \partial_{x_k} Q_{ij}\right)^2.\label{free}
 \end{equation}
In this formulation, $\dbar$ represents the variational derivative, ${\rm tr}$ denotes the trace, $K$ is an elastic constant, and $A$, $B$ and $C$ are material constants that determine the equilibrium state of the orientational order. The first three terms in (\ref{free}) correspond to the Landau/De Gennes free energy, while the last term represents the Frank elastic energy with a one-constant approximation \citep{DeGennesBook}. Equation~(\ref{free}) is the free energy expansion in terms of the order parameter, where the terms allowed by symmetry have been retained. Note that the Frank term in (\ref{free}) gives rise to a diffusive flux of $Q_{ij}$ in the form $\Gamma K \partial^{2}_{x_k,x_k}Q_{ij}$ in (\ref{eqn:lc}). The description of the flow field is completed by the mass and momentum conservation equations, namely
 \begin{equation}
\partial_{x_i}  u_{i}=0,\qquad \rho\partial_t u_i + \rho u_j \partial_{x_j}  u_i= -\partial_{x_i} p+\mu\partial^2_{x_j, x_j} u_{i}+\partial_{x_j} \sigma_{ij}-\zeta\partial_{x_j} Q_{ij},
 \label{eqn:ns}
 \end{equation}
 where $\rho$ is the density, $p$ is the pressure and $\mu$ is the dynamic viscosity.  The additional terms in the momentum equation (\ref{eqn:ns}) involve the elastic stress 
 \begin{align}
 \sigma_{ij}&= 2 \lambda\left(Q_{ij} + \delta_{ij}/3\right) (Q_{kl} H_{lk}) -\lambda H_{ik} \left(Q_{kj} + \delta_{kj}/3\right)  - \lambda \left(Q_{ik} + \delta_{ik}/3\right) H_{kj}\nonumber\\
 &-\partial_{x_i}  Q_{kl} \frac{\dbar \mathcal{F}}{\dbar \partial_{x_j} Q_{lk}} + Q_{ik}H_{kj} - H_{ik} Q_{kj},\label{eqn:st}
 \end{align}
which represents the passive conformational resistance of the nematics to changes in the orientational order \citep{Beris}, and the active stress $\zeta Q_{ij}$, which corresponds to a coarse-grained collective effect of the stresslets set up by the active particles \citep{Sriram2002}. In particular, the divergence of the active stress is responsible for the injection of kinetic energy through the particle alignment characteristics represented by the nematic tensor, whose conservation equation (\ref{eqn:lc}) is non-linearly coupled with the velocity field. The absolute value of the coefficient $\zeta$ is proportional to the activity, where positive and negative values of $\zeta$ correspond, respectively, to extensile and contractile particles, the former being the case addressed in the computations shown below. 
~\\[-0.25in]

 \subsection{Remarks on the conservation equations}\label{remarks}

Some aspects with regard to the formulation given above are worth briefly pointing out. The first one is related to characteristic dimensionless parameters. In conditions addressed by these simulations (i.e., see further details in~\S\ref{compu}) and most experimental observations, the Reynolds number is small, $Re=\rho u'\ell/\mu\lesssim 1$, with $u'$ and $\ell$ as the integral scales for velocity and length, respectively. In this limit, the convective transport of momentum has a diminishing influence, with the dominant balance in (\ref{eqn:ns}) being between viscous and active stresses. Similarly, the orientational order diffuses slower than momentum, as indicated by the relatively high Schmidt number $Sc=(\mu/\rho)/(\Gamma K)\gtrsim 5$, which results in characteristic structures of the velocity field that are larger than those of the orientational order, as discussed in \S\ref{qual}. As a consequence, the P\'{e}clet number of the nematic order $Pe=ReSc\gtrsim 3$ suggests that advection may play a more important role in transporting orientational order than it does for transporting momentum. 

The second aspect worth stressing is that the conservation equations outlined above do not include any obvious external forcing term aimed at sustaining the dynamics.  A flow laden with active matter is different from an inactive flow externally driven by boundary conditions, imposed shear or volumetric forces added to the momentum conservation equation. In real biological flows laden with swimming bacteria, adenosine tri-phosphate (ATP) molecules are consumed by the bacteria via hydrolysis reactions and transformed into motion in such a way that the driving occurs internally at the expense of depletion of chemical energy. The source of chemical energy, however, is not represented in the formulation above, in that it only concerns systems where the rate of ATP depletion is infinitesimally small compared to hydrodynamic processes. 

The active motorization of the flow can be understood by multiplying the momentum equation in (\ref{eqn:ns}) by $u_i$ and performing a surface (in 2D) or volumetric (in 3D) periodic integration, which leads to the balance equation
$\rho(dk/dt)=-\epsilon-\left\langle\sigma_{ij}S_{ij} \right\rangle+\zeta \left\langle Q_{ij}S_{ij} \right\rangle$ for the spatially averaged kinetic energy $k=\left\langle u_i u_i/2\right\rangle$, where $\epsilon=\left\langle 2\mu S_{ij}S_{ij} \right\rangle$ is the viscous dissipation. The work done by the active forces, given by the term $\zeta \left\langle Q_{ij}S_{ij} \right\rangle$, represents the main source of $k$. The power deployed by the active work is dissipated by viscosity as shown below, thereby yielding a stationary state in which $k$ remains mostly constant.

\subsection{Computational set-up}\label{compu}

The formulation described above is integrated numerically in quasi-2D and 3D domains with periodic boundary conditions. In the quasi-2D framework the velocity varies in 2D while the nematic directors could develop out of plane components, as in previous studies \citep{ourprl2013,ourNComm2016,Saw}. The dimensional parameters in the simulations are $A=0.04$, $B=0$ for 2D and $B=0.06$ for 3D, $C=0.06$, $\lambda=0.3$ (tumbling particles), $\Gamma=0.34$, $K=0.40$, $\rho=1.0$, and $\mu=0.66$ ($Sc=4.90$). The baseline activity coefficient is $\zeta_0=0.036$ for 2D and 3D, with an additional 2D computation being performed with a smaller activity $\zeta_0/10$. All parameters here are expressed in simulation units. This generic set of parameters has been shown in earlier work to reproduce qualitatively the active turbulent state observed in bacteria and microtubule/motor-protein suspensions \citep{Julia2012,Dogic2012,ourprl2013}. A standard lattice-Boltzmann approach is used to integrate (\ref{eqn:ns}), while (\ref{eqn:lc}) is solved by employing a second-order finite-difference predictor-corrector algorithm. The resulting set of ordinary differential equations is integrated in time using an Euler method. The number of grid points for the 2D and 3D simulations is $N=512^2$ and $128^{3}$, respectively, with a minimum grid spacing of $\Delta=\ell_d/3$ for all cases, with $\ell_d=\sqrt{K/A}=3.16$ being the characteristic size of the topological defects in the orientational order field. 
The time increment used in the numerical integrations is $\Delta t=t_q/60$, with $t_q= \mu/\zeta$ being a characteristic time scale for the damping of the activity by viscosity. The initial conditions consist of zero velocity everywhere while the directors are set to random orientations. Data are collected after approximately $2\times 10^{4}$ time steps for 10 snapshots spanning a time period of  $150t_q$, which provide $512^2\times 10$ and $128^3\times 10$ statistical samples leading to converged probability density functions (PDFs) in 2D and 3D, respectively. In the results, spatial coordinates are normalized with $\ell_d$. Additionally, velocities are normalized with $u'=0.11$ (for 2D, with $\zeta=\zeta_0$) and $u'=0.04$ (for 3D, and 2D with $\zeta=\zeta_0/10$), where $u'=\sqrt{2k}$. 
~\\[-0.3in]
\section{Analysis of numerical results}\label{qual}
 
\subsection{Flow structures}
\begin{figure}
	\begin{center}
	    	\includegraphics[width = 0.8\textwidth]{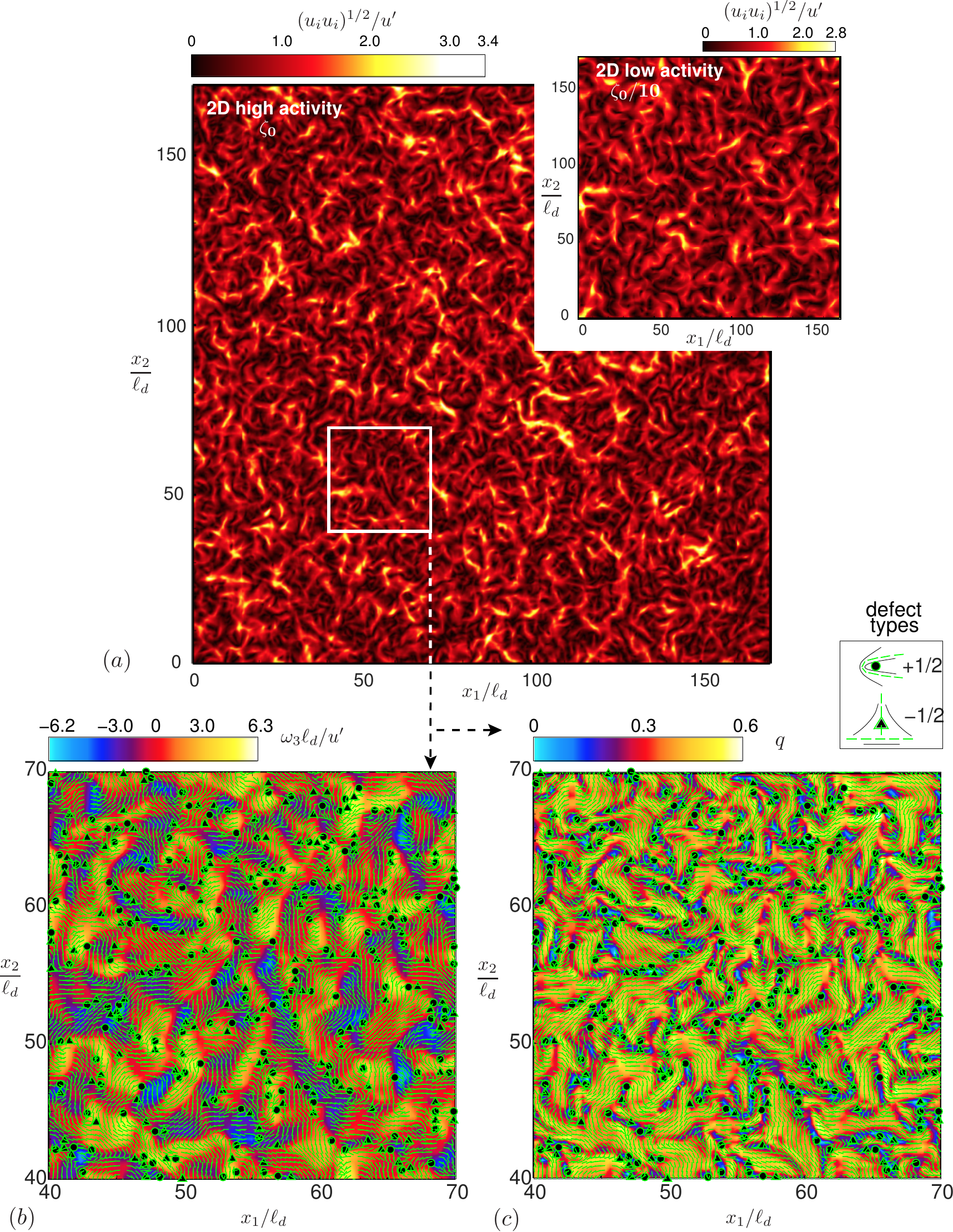}
      	\caption{Instantaneous contours of (a) velocity magnitude (high activity, main panel; low activity, inset), (b) vorticity and (c) magnitude of the nematic tensor. In panels (b,c), which are zoomed views of the white squared region in panel (a), green lines are nematic director fields, while symbols represent $+1/2$ (circles) and $-1/2$ (triangles) topological defects. The small inset
above (c) shows the director field around topological defects. \label{Fig1}}
	\end{center}
			\vskip -0.18in

\end{figure}

\begin{figure}
	\begin{center}
	\vskip 0.1in
		    	\includegraphics[width = 0.7\textwidth]{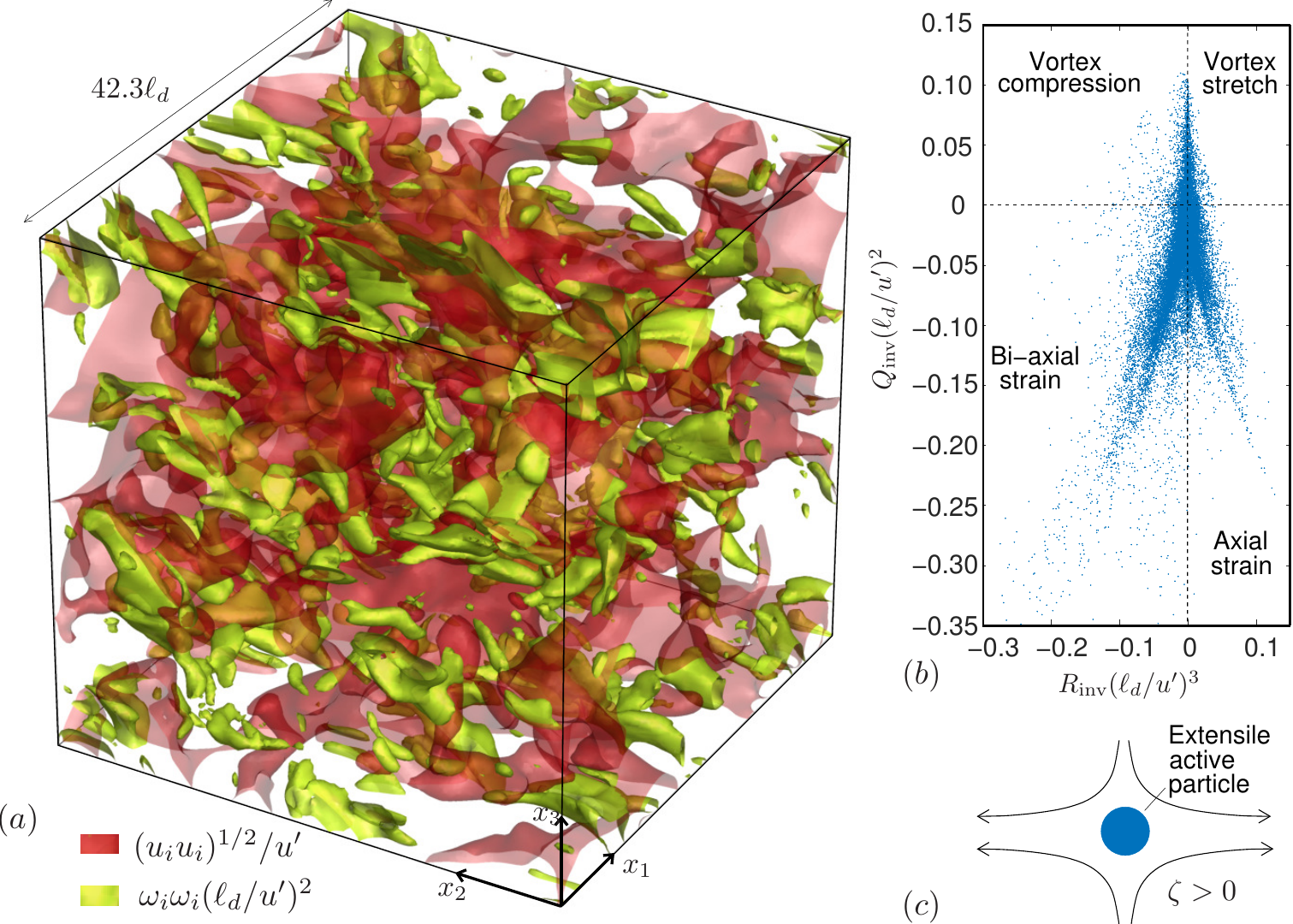}
      	\caption{(a) Instantaneous iso-surfaces of velocity magnitude $(u_iu_i)^{1/2}/u'=1.1$ (30\% of maximum value) and enstrophy $\omega_i\omega_i(\ell_d/u')^2=0.4$ (20\% of maximum value) for 3D simulations. (b) Scatter plot of the velocity-gradient invariants. (c) Schematics of the stresslet-like flow induced by extensile active particles ($\zeta>0$).  \label{Fig2}}

	\end{center}
\end{figure}

Instantaneous contours of velocity and vorticity are provided in figures~\ref{Fig1}(a),(b) and figure~\ref{Fig2}(a) for 2D and 3D domains, respectively. Specifically, the effects of decreasing the activity coefficient from $\zeta_0$ to $\zeta_0/10$ in the 2D simulations are an increase in the integral length (computed from the ensemble-averaged kinetic-energy spectrum) from $\ell=5.45$ to $\ell=10.22$, a decrease in the Reynolds number from to $Re=0.74$ to $Re=0.68$, and a decrease in the dissipation from $\epsilon=1\cdot 10^{-3}$ to $\epsilon=2\cdot 10^{-4}$. The resulting low-activity flow has a less dense pattern of flow structures, as observed by comparing the main (high activity) and inset (low activity) frames in figure~\ref{Fig1}(a) and noticing that the normalizing length $\ell_d$ is independent of activity. It is noteworthy that, for the same activity coefficient $\zeta_0$, moving from two to three dimensions translates into an increase in the integral length from $\ell=5.45$ to $\ell=8.97$, a decrease in the Reynolds number from $Re=0.74$ to $Re=0.53$, and a decrease in the dissipation from $\epsilon=1\cdot 10^{-3}$ to $\epsilon=0.3\cdot 10^{-4}$. 

The spatial variations in the nematic tensor $Q_{ij}$ are central to the generation of vorticity. This is easily observed by taking the curl of (\ref{eqn:ns}), namely
\begin{equation}
\rho\partial_t \omega_i+\rho u_j\partial_{x_j} \omega_i=-\rho \omega_j\partial_{x_j} u_i +\varepsilon_{ijk}\partial_{x_j}\left(\mu\partial^{2}_{x_{\ell},x_{\ell}} u_{k}+\partial_{x_{\ell}}\sigma_{k\ell}-\zeta\partial_{x_{\ell}} Q_{k\ell}\right),\label{vort}
\end{equation}
where $\omega_i$ is the vorticity and $\varepsilon_{ijk}$ is the permutation tensor. In 2D, the vortex stretch term is exactly zero and the dominant mechanism of vorticity generation is the curl of the divergence of the active stresses. The structures of vorticity, which are shown in figure~\ref{Fig1}(b), are different from the classic round vortices observed in high-Reynolds number 2D isotropic flows. Instead, vortical structures attain here band-like shapes, which are closely related to thinner elongated regions referred to as walls, where the magnitude of the nematic order tensor becomes small, as shown by the solid contours of $\omega_3$ overlaid on the director field (largest eigenvector of the $Q_{ij}$ tensor) in figure~\ref{Fig1}(b). The walls are characterized by bend deformations in the in-plane director field in figure~\ref{Fig1}(c) (note that the out-of-plane components in these 2D simulations are zero), which separate nematically-aligned regions ($q\sim 1$)  across interstitial isotropic states ($q\ll 1$), and are typically much thinner than the hydrodynamic structures of velocity and vorticity. The resulting $\pm 1/2$ topological defects, depicted by circles and triangles in figure~\ref{Fig1}(b,c) (see inset), represent singular, disordered regions of strong vorticity generation that are created and annihilate in pairs while propagating in the flow in a complex manner \citep{prl} that is beyond the scope of the present study. 

In 3D, the vortex-stretch term in (\ref{vort}) represents a much smaller contribution than the active stresses because of the low Reynolds numbers involved. As observed in figure~\ref{Fig2}(a), the resulting vortical structures in 3D are elongated as well but smaller than the velocity ones. The 3D mechanisms of vorticity generation are mostly unknown since the description of topological defects in active nematics is not well understood. Further insight into rotational and straining components of the 3D flow field can be gained by examining the velocity-gradient invariants $Q_{\textrm{inv}}=(1/4)(\omega_i\omega_i-2S_{ij}S_{ij})$ and $R_{\textrm{inv}}=(3/4)(\omega_i\omega_jS_{ij}+4S_{ij}S_{jk}S_{ki})$, which are expedient to classify flow structures. In contrast to typical scatter plots for high-Reynolds turbulence where most of the activity is in the upper-right and lower-left quadrants, figure~\ref{Fig2}(b) shows that for active flows straining is predominant. As a result, the marginal PDF of $Q_{\textrm{inv}}$ is heavily skewed toward negative values (skewness $-1.6$). The straining is caused by the cumulative effect of the stresslets from the extensile active particles (i.e., see flow sketch in figure~\ref{Fig2}(c)).  
\begin{figure}
	\begin{center}
	    	\includegraphics[width = 0.9\textwidth]{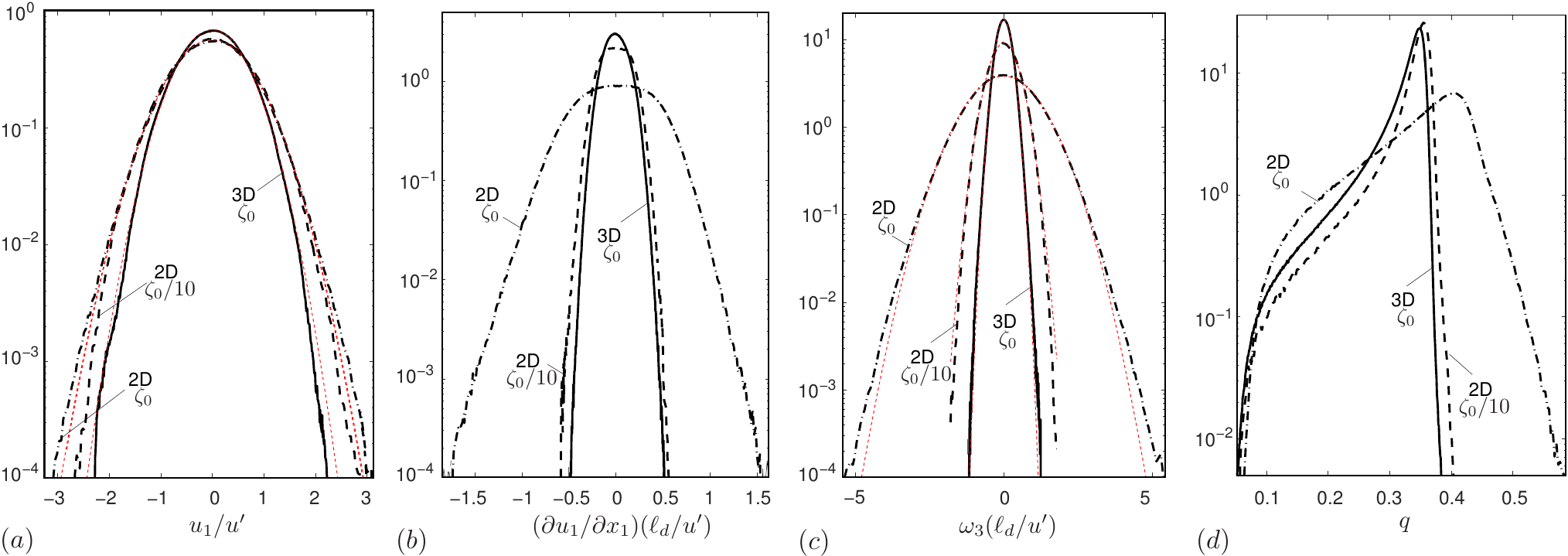}
      	\caption{Ensemble-averaged PDFs of (a) velocity, (b) velocity gradient, (c) vorticity and (d) nematic-order magnitude, including 3D ($\zeta=\zeta_0$, solid lines) and 2D ($\zeta=\zeta_0$, dot-dashed lines; $\zeta=\zeta_0/10$, dashed lines). Red short-dashed lines indicate reference Gaussian distributions.\label{Fig4}}
	\end{center}
\end{figure}
~\\[-0.2in]
\subsection{PDF moments of flow variables}
The PDFs of velocity $u_1$, velocity gradient $\partial u_1/\partial x_1$, vorticity $\omega_3$ and magnitude of the nematic order $q$ are shown in figure~\ref{Fig4}, and some of their  moments are listed in Table~\ref{table1}. The PDFs of velocity and vorticity have nearly Gaussian flatness, while the largest skewness of the velocity gradient is reached in the 2D high-activity case and equals $-0.10$.  In all cases, the vorticity flatness and the velocity-gradient skewness are smaller than typical values observed in high-Reynolds turbulence (i.e., $\sim 8$ and $\sim -0.4$, respectively). In 2D, small activities favor sub-Gaussian flatness for vorticity and velocity along with decreasing skewness of the velocity gradient. Note, however, that for the same activity coefficient the 3D case leads to a smaller skewness of the velocity gradient, which suggests that the lesser spatial confinement plays a role in the development of fluctuations. Additionally, small values of the nematic-order magnitude, which correspond to isotropic behavior and vorticity generation, are statistically favored by large activities. 
\begin{table}
\begin{center}\scriptsize
\begin{tabular}{lccc}
&&&\\[-5ex]
 & 2D ($\zeta_0/10$) & 2D ($\zeta_0$) & 3D ($\zeta_0$)\\[1ex]
Velocity ($u_1$) flatness & 2.84 & 3.14 & 2.78 \\
 Velocity derivative ($\partial u_1/\partial x_1$) skewness &-0.03 & -0.10 &  -0.02 \\
 Vorticity ($\omega_3$) flatness & 2.85 & 3.21 &  3.12 \\
\end{tabular}
\caption{Moments of velocity, velocity derivative and vorticity PDFs. \label{table1}}
\end{center}
\end{table}

\subsection{Intermittency analysis}

\begin{figure}
	\begin{center}
	    	\includegraphics[width = 0.8\textwidth]{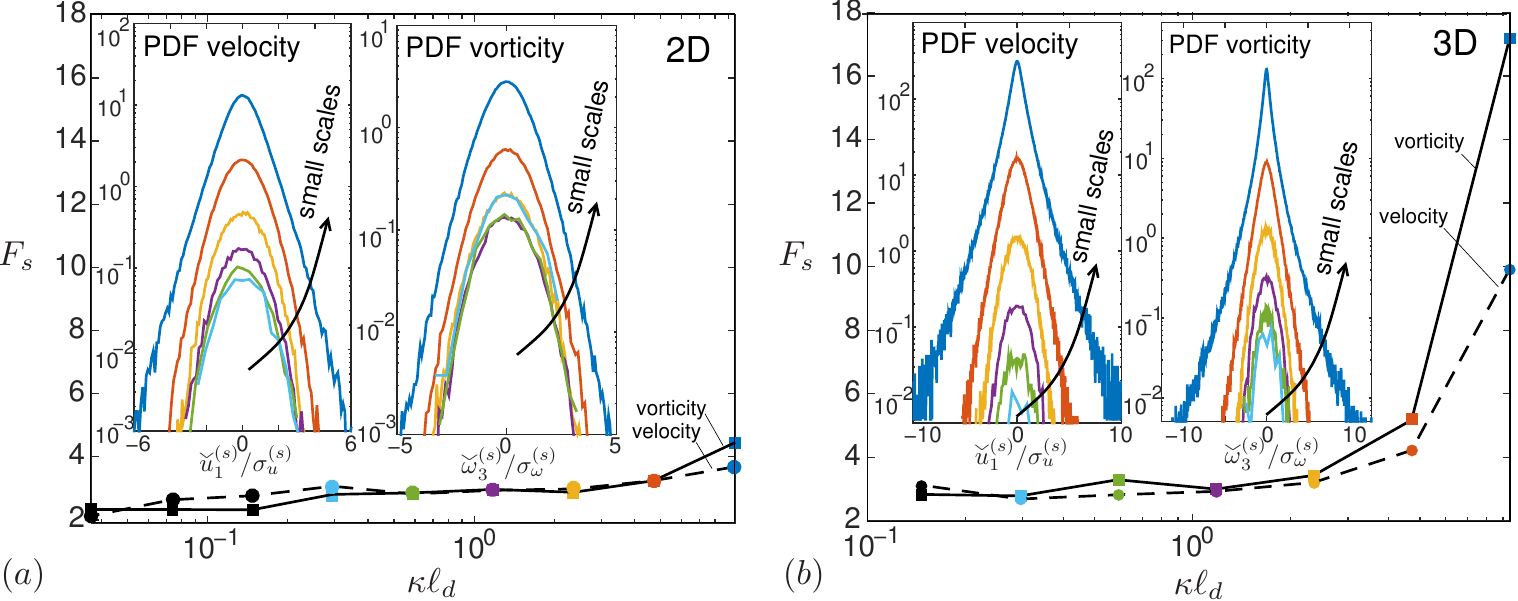}
      	\caption{Wavelet-based scale-dependent flatness (main frames) for velocity and vorticity in (a) 2D and (b) 3D cases at $\zeta=\zeta_0$, along with corresponding scale-conditioned PDFs (insets) of the wavelet coefficients. The PDFs are conditioned on $s=1$ (dark blue lines) $s=2$ (red), $s=3$ (orange), $s=4$ (purple), $s=5$ (green) and $s=6$ (light blue).\label{Fig5}}
	\end{center}
\end{figure}

Albeit small, intermittency may not be entirely ruled out in these systems, as suggested by a narrow-band filtering analysis of the results based on a discrete db-4 wavelet decomposition of the velocity and vorticity fields. This is shown in figure~\ref{Fig5}, which provides the scale-dependent flatness $F^{(s)}$ of the PDFs of the direction-averaged velocity and vorticity wavelet coefficients, $\widecheck{u}^{(s)}_1$ and $\widecheck{\omega}^{(s)}_3$, normalized with their corresponding standard deviations $\sigma_{u}^{(s)}$ and $\sigma_\omega^{(s)}$, with $s$ being a scale index that ranges from $1$ (equivalent to a length scale $2\Delta$) to $s_{\textrm{max}}=\log_2 N$ (equivalent to $N\Delta$) and is related to a representative wavenumber as $\kappa=2\pi 2^{-s}/\Delta$. In these simulations, $s_{\textrm{max}}=9$ (for 2D cases) and $s_{\textrm{max}}=7$ (for 3D cases). The inverted caret denotes the wavelet transform, e.g. $\widecheck{u}_i^{(s,d)}(\bxs)=\left\langle u_i(\boldsymbol{x})\Psi^{(s,d)}(\boldsymbol{x}-\boldsymbol{x_s})\right\rangle$, where $d$ is a positive-integer direction index ($d=1,2,3$ and $d=1,2...,7$ in 2D and 3D, respectively), $\boldsymbol{x_s}=2^{s-1}(i\Delta,j\Delta,k\Delta)$ are scale-dependent wavelet grids where the wavelets are collocated, with $i$, $j$, $k=1,3,5\dots,$ $N/2^{s-1}-1$, and $\Psi^{(s,d)}(\boldsymbol{x}-\boldsymbol{x_s})$ are wavelet basis functions that are here taken to be tensor products of orthonormal one-dimensional db-4 wavelets with four vanishing moments. The reader is referred to \cite{Meneveau} and \cite{Schneider} for general applications of wavelets in turbulent flows, and to \cite{Farge} for a scale-dependent flatness analysis of homogeneous isotropic turbulence similar to the one performed here.

\begin{figure}
	\begin{center}
	    	\includegraphics[width = 0.8\textwidth]{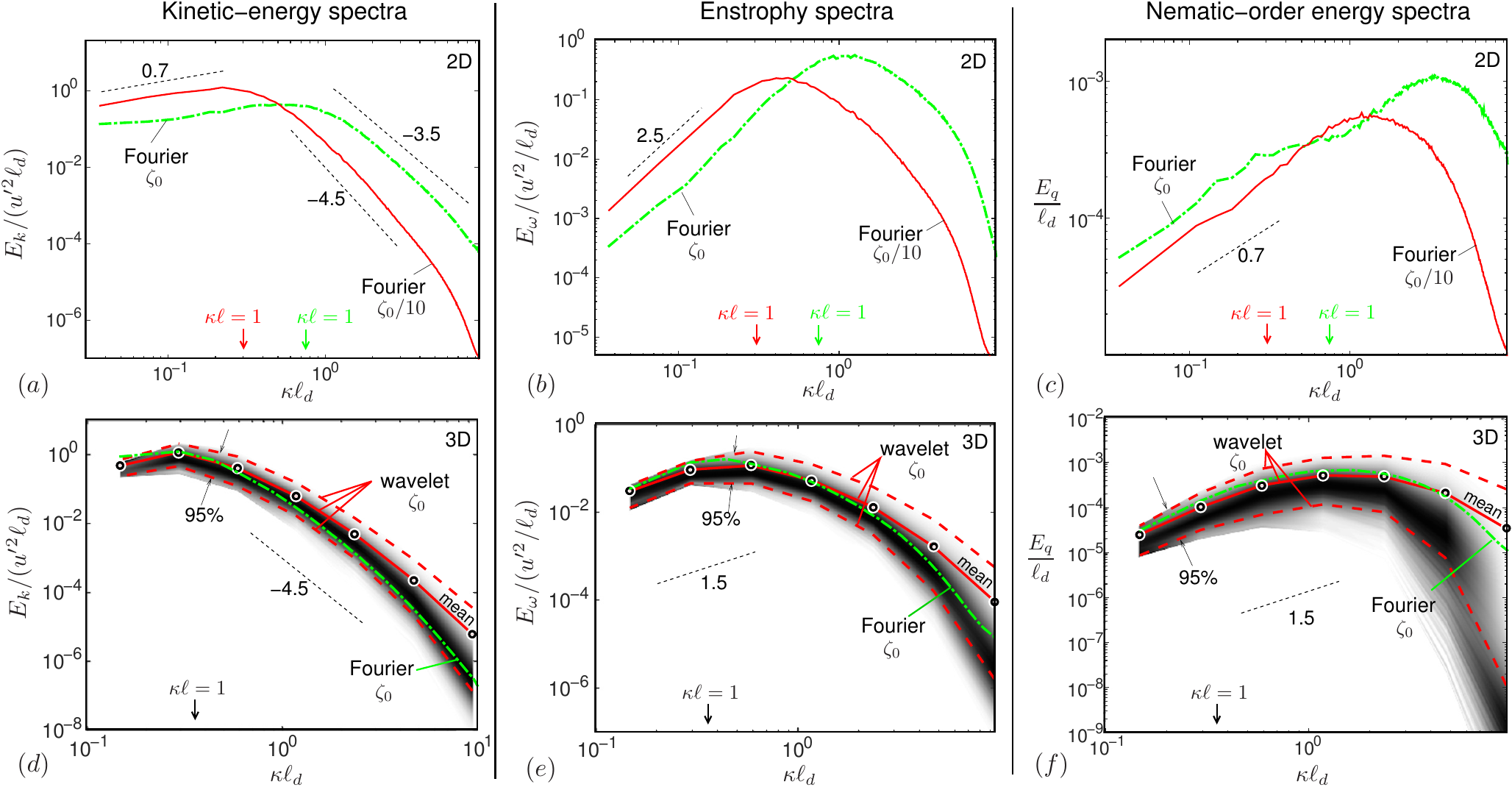}
      	\caption{Ensemble-averaged Fourier and wavelet spectra of kinetic energy. The solid contours in the bottom row correspond to the PDF of the wavelet spectra, which include the mean (solid lines) and associated 95\% confidence intervals (dashed lines).\label{Fig7}}
	\end{center}
\end{figure}

While the 2D fields remain nearly Gaussian at all scales, with a slight increase for the vorticity flatness observed in the largest scale, figure~\ref{Fig5}(b) shows that the  3D fields contain significant intermittency in the small scales, as indicated by the strong increase in $F_s$ with $\kappa$ (main panel) and by the increasingly longer tails in the PDFs of the wavelet coefficients as the length scale increases (inset). Nonetheless, the energetic content of these small scales and their associated intermittent motion is small in all cases. This can be understood by noticing the rapid decay of the Fourier kinetic-energy and enstrophy spectra, $E_k$ and $E_\omega$, as $\kappa$ increases, as observed in figure~\ref{Fig7}(a,b,d,e). Specifically, the kinetic-energy spectra decay with a slope that decreases from $-4.5$ to $-3.5$ as the activity is increased ten-fold. As a result, the enstrophy spectra reach a maximum at the integral scale and decay rapidly thereafter, indicating that the small-scale gradients bear vanishing energy. This detracts dynamical relevance from the increased intermittency observed in the 3D small scales and sets a fundamental difference between these flows and high-Reynolds number turbulence; in the latter, the kinetic-energy spectra decays at a slower rate and the small-scale vorticity intermittency is energetic. 

It is also of interest to note that the characteristic wavenumber where the nematic-order fluctuation energy spectra $E_q$ (defined such that the area under the curve is $\langle q'q'\rangle$) reach a maximum is approximately one decade smaller than the wavenumber corresponding to the maximum enstrophy spectra in the 2D low-activity case, as shown in figure~\ref{Fig7}(c). The distance between peaks decreases with increasing activity and when moving to three dimensions, as observed in figure~\ref{Fig7}(f). The wavenumber of maximum $E_q$ decreases as the activity increases and its value is closer to $2\pi/\ell_d$ than to $2\pi/\ell$, which spectrally illustrates the fine structure of the nematic-order field compared to the coarser velocity field.

\begin{figure}
	\begin{center}

	    	\includegraphics[width = 0.8\textwidth]{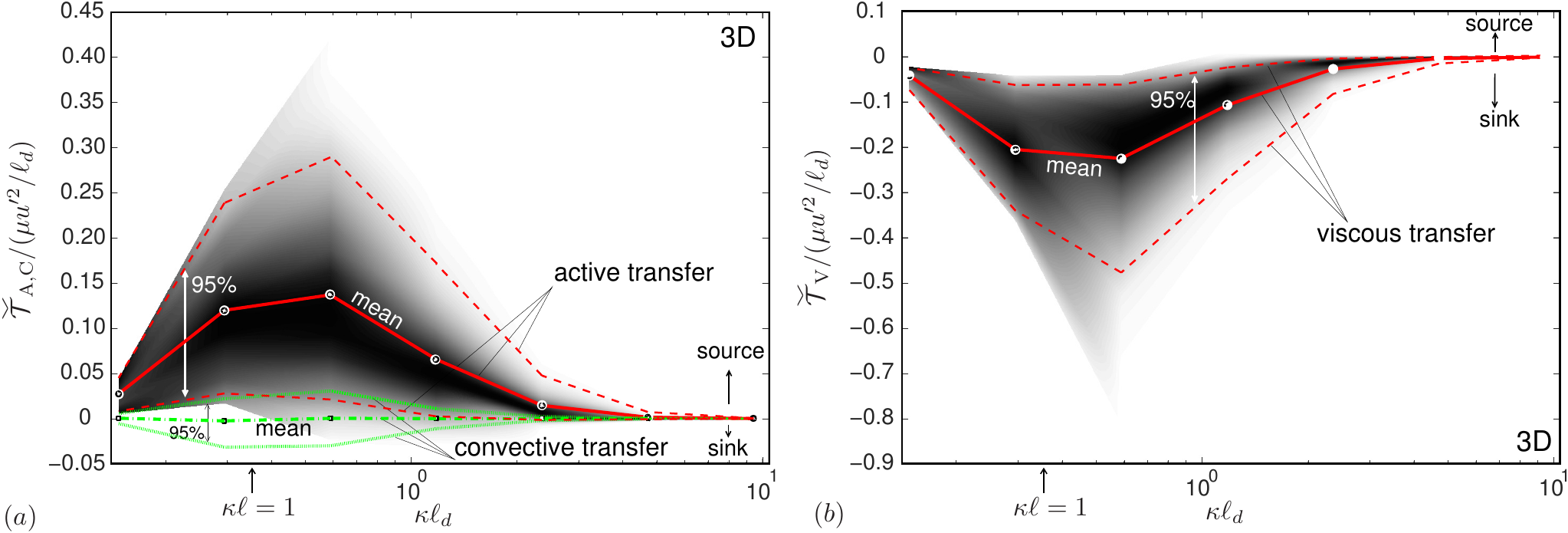}
      	\caption{Mean and 95\% confidence intervals of wavelet-based spectral energy-transfer flux of the 3D flow for (a) active and convective fluxes, and (b) viscous flux. The panels also show solid contours for PDFs of (a) active and (b) viscous fluxes, indicating spatial variabilities.\label{Fig8}}
		\end{center}
\vskip -0.15in
\end{figure}

\subsection{Spectral energy-transfer analysis}
As discussed in \S\ref{remarks}, a crucial role in the dynamics is played by the divergence of the active stress $-\zeta Q_{ij}$. Because of the small $Re$ involved, it is anticipated that the spectral transfer of the active energy is locally dissipated by viscosity, since the convective inter-scale transfer is a mechanism of secondary importance. Although there may exist additional triadic interactions resulting from (\ref{eqn:st}) that could transport energy across scales, the 3D results provided in figure~\ref{Fig8} for active ($\widecheck{\mathcal{T}}_{\textrm{A}}$), convective ($\widecheck{\mathcal{T}}_{\textrm{C}}$), and viscous ($\widecheck{\mathcal{T}}_{\textrm{V}}$) wavelet-based spectral energy-transfer fluxes support the view that locality may dominate the transfer. Specifically, these fluxes describe the rate at which the spatially averaged, wavelet spectral kinetic-energy density, $E_k=2^{-3s}\langle\sum_{d}\widecheck{u}_i^{(s,d)}(\bxs)\widecheck{u}_i^{(s,d)}(\bxs)/2\rangle_{\bxs}/\delta\kappa$, is transferred across scales. In particular, $E_k$ and the analogous $E_\omega$ and $E_q$  are shown and compared to their Fourier-based counterparts in figure~\ref{Fig7}(d-f). In this formulation, $\delta\kappa= 2\pi\ln 2/(2^s\Delta)$ is a discrete wavenumber shell, and the subindexed bracketed operator represents spatial averaging over scale-dependent wavelet grids $\boldsymbol{x_s}$. 

Upon wavelet-transforming the momentum equation in (\ref{eqn:ns}), multiplying by $\widecheck{u}_i^{(s,d)}(\bxs)$ and summing over $d$, the spectral-energy equation $\partial E_k/\partial t=\sum \widecheck{\mathcal{T}}(\kappa)$ is obtained. Here, the source term represents the sum of spectral energy-transfer fluxes created by each term on the right-hand side of the momentum equation in (\ref{eqn:ns}). For any force $\phi_i$, the corresponding spectral flux is given by $\widecheck{\mathcal{T}}(\kappa)=[(2^{-2s}\Delta)/(2\pi\ln 2)]\langle \sum_{d}\widecheck{u}_i^{(s,d)}(\bxs)\widecheck{\phi}_i^{(s,d)}(\bxs)\rangle_{\bxs}$, with $\widecheck{\mathcal{T}}>0$ and $\widecheck{\mathcal{T}}<0$ indicating, respectively, inflow and outflow of energy at a given $\delta\kappa$. Note that $\widecheck{\mathcal{T}}=\widecheck{\mathcal{T}}_{\textrm{A}}$ for $\phi_i=-\zeta\partial_{x_j}Q_{ij}$, $\widecheck{\mathcal{T}}=\widecheck{\mathcal{T}}_{\textrm{V}}$ for $\phi_i=\mu\partial^{2}_{x_j,x_j}u_{i}$, and $\widecheck{\mathcal{T}}=\widecheck{\mathcal{T}}_{\textrm{C}}$ for $\phi_i$ $=-\rho u_j\partial_{x_j}u_i$, which satisfy $\sum_{\kappa} \widecheck{\mathcal{T}}_{\textrm{A}}\delta\kappa=\zeta\langle Q_{ij}S_{ij}\rangle$, $\sum_{\kappa}  \widecheck{\mathcal{T}}_{\textrm{V}}\delta\kappa=-\epsilon$ and $\sum_{\kappa} \widecheck{\mathcal{T}}_{\textrm{C}}\delta\kappa=0$. 

Figure~\ref{Fig8}(a) indicates that the active stress act as a kinetic-energy source at all scales on spatial average, with the maximum mean of $\widecheck{\mathcal{T}}_{\textrm{A}}$ occurring at scales of the same order as the integral length. Conversely, the viscous flux $\widecheck{\mathcal{T}}_{\textrm{V }}$ is a sink of kinetic energy and has a trend that is exactly opposite to $\widecheck{\mathcal{T}}_{\textrm{A}}$, as shown in Figure~\ref{Fig8}(b), which indicates that the active energy is mostly dissipated locally in spectral space by viscosity. The spatial localization of the transfer, which is illustrated by the unbracketed versions of $E_k$, $E_\omega$, $E_q$ and $\widecheck{\mathcal{T}}$, is represented by the variabilities of the PDFs shown in figures~\ref{Fig7}(d-f) and \ref{Fig8}. Specifically, the PDFs in figure~\ref{Fig8} reveal that the viscous transfer flux is spatially correlated with the active one (correlation coefficient $-0.73$ at $s=4$), suggesting that upon deployment the active energy is dissipated mostly locally also in physical space.

The physical picture implied by figure~\ref{Fig8} provides no evidence for an energy cascade in momentum where the sink $\epsilon$ and main source $\zeta\langle Q_{ij}S_{ij}\rangle$ of mean kinetic energy could act in disparate ranges of scales interacting through a crossing long-range mechanism. This is in contrast to high-Reynolds number turbulence and its clear separation of scales between the large-scale forcing range and the small-scale molecular-dissipation range. These conclusions could however be different for the nematic-order energy, in that the transport description of the latter is highly non-linear and involves cross-triadic terms with the velocity as in (\ref{eqn:cor}). These aspects will be the subject of future research.

\section{\label{conclusion}Conclusions}
The multi-scale statistical analysis of DNS presented in this study provides quantitative comparisons between active and classic turbulent flows beyond superficial visual similarities, demonstrating clear distinctions in the intermittency characteristics and mechanisms of inter-scale energy transfer. It is shown that increasing activities lead to increasingly packed and dissipating structures that have increasingly larger departures from Gaussian statistics. For the same activity, the 3D flow has a larger integral length and smaller kinetic energy compared to its 2D counterpart. A velocity-gradient invariant analysis of the 3D flow indicates that straining structures dominate the topology as a collective result of the embedded stresslets induced by each individual extensile active particle. A wavelet-based, scale-dependent flatness analysis shows the occurrence of intermittency in the small scales, particularly in the 3D vorticity field. However, the spectral energy content associated with the small-scale velocity gradients is small in all cases. The work of the active stress is spectrally deployed near the integral scales and mostly dissipated locally by viscosity in both physical and spectral spaces. 

\subsection*{Acknowledgments}
This investigation was performed during the 2016 CTR Summer Program at Stanford University. The authors are grateful to Maxime Bassenne, Dr. Jeonglae Kim, and Profs. Marie Farge and Kai Schneider for useful discussions.
\bibliographystyle{ctr}

\end{document}